\documentclass[aps,pre,onecolumn,nofootinbib,showkeys,longbibliography]{revtex4-2}
\usepackage{CJKutf8}          
\usepackage{amsmath,amssymb,bm}
\usepackage{graphicx}
\usepackage{color,soul}
\usepackage[colorlinks=true,linkcolor=blue,citecolor=blue,urlcolor=blue]{hyperref}
\usepackage{orcidlink}        
\usepackage{changes}
\def\la{\langle}
\def\ra{\rangle}
\newcommand{\imu}{\mathrm{i}}
\newcommand{\dd}{\mathrm{d}}
\newcommand{\eff}{\mathrm{eff}}
\newcommand{\qs}{\mathrm{qs}}
\newcommand{\fric}{\mathrm{fric}}

\usepackage[most]{tcolorbox} 
\begin{document}

\title{Energy transport from an active medium}

\author{Simon Krekels\,\orcidlink{0000-0003-1726-6625}}
\author{Christian Maes\,\orcidlink{0000-0002-0188-697X}}
\author{Ion Santra\,\orcidlink{0000-0002-9772-2880}}
\affiliation{Department of Physics and Astronomy, KU Leuven, Celestijnenlaan 200D, 3001 Leuven, Belgium}

\date{\today}

\begin{abstract}
We study the energy current through a harmonic chain whose left end is coupled, via a short-range contact potential, to a medium of overdamped run-and-tumble particles moving on a ring, while the right end is attached to a zero-temperature Langevin reservoir.
The activity is parametrized by the propulsion speed $v_0$ and the persistence time $\tau$.  
Using nonequilibrium response theory at weak system-medium coupling, we predict the stationary current and simulations reproduce it over a large part of the $(v_0,\tau)$ plane.
That includes the possible nonmonotonicity of the current as function of $v_0$ and $\tau$, and occurrence of resonances with the normal modes of the chain, most significantly at large $\tau$ and small $v_0$.  There is a critical propulsion speed $v_c$ at which the ring-crossing frequency of the particles leaves the phonon band: for $v_0<v_c$ the current grows with
persistence, while for $v_0>v_c$ it decays as $(v_0^2\tau)^{-1}$.
We also explain where the theoretical prediction fails, especially due to an oscillatory instability originating from a sufficiently strong negative friction induced on the leftmost oscillator.
Finally, the persistent medium produces strongly non-Gaussian velocity statistics and long-range, oscillating velocity correlations in the bulk of the chain.  
\end{abstract}

\keywords{active matter; run-and-tumble particles; heat conduction; harmonic chain; active reservoirs; memory friction; negative friction; negative differential conductance}

\maketitle

\section{Introduction}\label{sec:intro}

Heat conduction through a chain of oscillators placed between two reservoirs is a standard testing ground of nonequilibrium statistical mechanics~\cite{RiederLebowitzLieb, LepriLiviPoliti, DharReview}.
For harmonic chains, the problem is exactly solvable: transport is ballistic, the bulk temperature profile is flat, and the stationary current takes the Landauer--B\"uttiker form, an integral over the phonon band of a transmission coefficient times the spectral difference of the reservoirs~\cite{ DharRoy, CasherLebowitz}.
In that setting the reservoirs are equilibrium Langevin baths, fully characterized by a temperature and a friction coefficient related through the Einstein relation.  However, when the contact kinetics makes the friction temperature-dependent, a negative differential conductivity may result at high temperature-differences~\cite{ndc2026}. \\

The present paper replaces one reservoir by a nonequilibrium medium. Active media, composed of units that consume energy to propel themselves, are a prominent example~\cite{Marchetti2013,Bechinger2016}. They are not characterized by a temperature, their fluctuations are \emph{colored}, and the friction and noise they exert on an immersed probe are no longer tied by the fluctuation--dissipation theorem~\cite{MaesSteffenoni2015,ActiveFriction,santra2023dynamical}. Probes in such media show enhanced diffusion~\cite{WuLibchaber2000,Maggi2014,Granek2022}, experience nonequilibrium induced forces~\cite{Basu2015forces,santra2024forces}, relax in a way that breaks time-reversal symmetry~\cite{KnotzKruger2024MBR}, and can even feel a negative friction, so that the medium feeds energy into their motion~\cite{pei2025induced,pei2026transfer,Woo2026trajectory}.\\  
Attempts at understanding activity-driven energy transport have so far focused on exactly solvable settings, namely, harmonic chains driven by active forces at their boundaries~\cite{santra2022activity,sarkar2023stationary}, and chains coupled to linear active mediums of oscillators~\cite{sarkar2024rubin}. These models already show nontrivial transport, including a negative differential conductivity and a reversal of the current with increasing activity~\cite{santra2022activity,sarkar2024rubin}. Owing to the linearity of the coupling, however, activity enters there only through noise spectrum of the active reservoir, while the induced friction remains that of a passive bath.  However, coupling with a soft matter system may be far from linear. Colloids or microswimmers interact with an immersed object locally through short-range forces, so that the object is kicked only when particles pass close to it, and the particles are in turn deflected by the object. Such a coupling is studied here, where an active medium is short-range coupled to a conductor.\\
The first particle of the conductor interacts with independent run-and-tumble particles (RTP) on a ring through a localized contact potential. The activity of the medium is characterized by the propulsion speed $v_0$ and the persistence time $\tau$ of the particles. Similar interactions have been studied for an elastic string and an oscillator chain coupled to an active medium~\cite{beyen2025coupling,beyen2026agitated}. To isolate the effects of activity on transport, as conductor we keep the harmonic chain, and the reservoir at its other end remains a Langevin bath at zero temperature.\\

Assuming a weak system-medium coupling, with a separation of time scales between the active medium and the chain as a sufficient condition, we use nonequilibrium response theory to derive a reduced non-Markovian linear dynamics for the chain.  We find that the active medium induces a negative stiffness, a memory friction that can become negative, and a colored noise on the first oscillator. The current of the reduced linear system reproduces the simulated current over a large part of the $(v_0,\tau)$ plane, including an interesting resonance-like phenomena with the normal modes of the chain. The current changes character at the speed $v_c$ at which the ring-crossing frequency of the active particles  leaves the phonon band. Above $v_c$, the current decays as $\tau^{-1}$ at large persistence
and as $v_0^{-2}$ at large propulsion speed. Below $v_c$, it grows
with persistence and reaches its largest values.
We also find that the kinetic temperature profile of the chain is flat in the bulk, and its bulk value follows the same trends as the current in the $(v_0,\tau)$ plane. The higher order velocity correlations, in contrast, reveal a strong imprint of the active medium. The velocity distributions may become skewed at the two ends and multi-peaked in the bulk, and the velocity of the oscillators also exhibits non-trivial spatial correlations.\\

The outline of the paper is as follows. Section~\ref{sec:model} introduces the model. Section~\ref{sec:probe} derives the reduced dynamics of a single probe in the active medium: induced stiffness, memory friction and noise. Section~\ref{sec:current} applies it to the chain: the Landauer--B\"uttiker formula and the two types of instability (Sec.~\ref{sec:instab}), the current as a function of $v_0$ including the transition at $v_c$ and the negative differential conductance (Sec.~\ref{sec:v0}), and as a function of $\tau$ together with the modified theory in the region of the first instability (Sec.~\ref{sec:tau}).
Section~\ref{sec:bulk} discusses temperature profiles, velocity distributions and spatial correlations in the bulk, and Sec.~\ref{sec:concl} concludes.

\section{Model}\label{sec:model}
\begin{figure}
    \centering
    \includegraphics[width=0.75\linewidth]{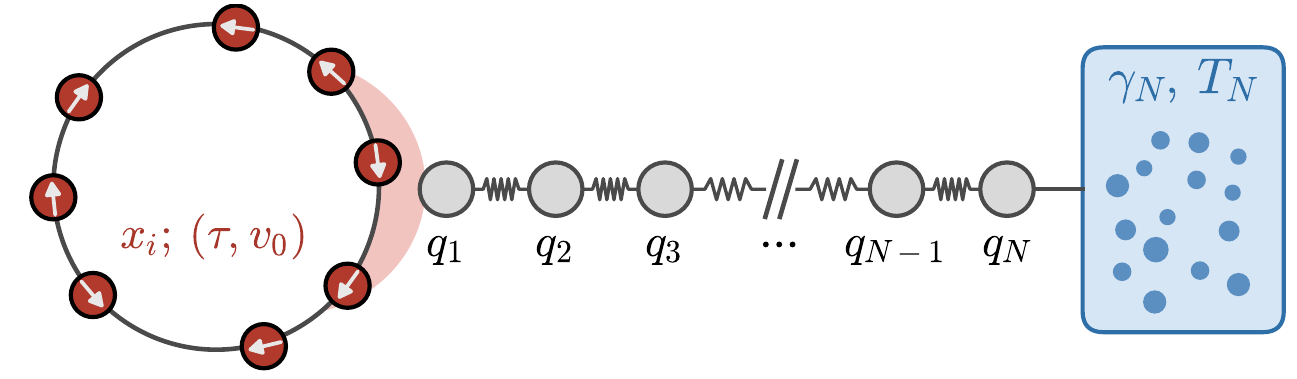}
\caption{Schematic of the setup. A harmonic chain of $N$ oscillators positions $q_1,\dots,q_N$ is coupled at its left end to an active contact medium of run-and-tumble particles (RTPs) on a ring, with positions $x_i$, persistence time $\tau$ and propulsion speed $v_0$. The right end of the chain is coupled to a thermal Langevin bath with friction coefficient $\gamma_N$ and temperature $T_N$.}
    \label{setup}
\end{figure}
The conductor is a one-dimensional chain of $N$ coupled harmonic oscillators with mass $m$ positions $\{q_j\}$ and momenta $\{p_j\}$, $j=1,\ldots,N$.  The first oscillator is coupled to a reservoir of $N_b$ overdamped active particles with positions $\{x_i\}$ on a ring of length $L$, and its last oscillator feels a Langevin reservoir. The chain and its contact with the medium are described by the Hamiltonian,
\begin{align}
     \mathcal H=\sum_{j=1}^N\frac{p_j^2}{2m}+\frac k2\sum_{j=1}^{N-1}(q_{j+1}-q_j)^2+\frac K2\left(q_1^2+q_N^2\right)-\Lambda  q_1
    +\lambda q_1\sum_{i=1}^{N_b}\mathcal F(x_i).
    \label{eq:hamiltonian}
\end{align}
Here $k>0$ is the chain stiffness, and $K>0$ are wall pinnings, and $\Lambda>0$,  which we specify later. The last term  characterizes the system-medium interaction with coupling strength $\lambda>0$. Figure~\ref{setup} shows a schematic for the setup. The equations of motion corresponding to the above Hamiltonian with the Langevin reservoir, characterized by $\gamma_N,$ $T_N$, at the last particle are,
\begin{subequations}\label{eq:chain-all}
\begin{align}
     m\ddot q_1
     &=
     -Kq_1-k(q_1-q_2)
     -\lambda\sum_{i=1}^{N_b}\mathcal F(x_i)+\Lambda,
     \label{eq:eom-q1}\\
     m\ddot q_j
     &=
     k(q_{j-1}-2q_j+q_{j+1}),
     \qquad j=2,\ldots,N-1,
     \label{eq:eom-bulk}\\
     m\ddot q_N
     &=
     -Kq_N-\gamma_N\dot q_N-k(q_N-q_{N-1})
     +\sqrt{2\gamma_N T_N}\,\eta_N,
     \label{eq:eom-qN}
\end{align}\label{eq:eoms}
\end{subequations}
where $\eta_N$ is standard white noise.
The medium particles with positions $x_i$ follow overdamped run-and-tumble dynamics~\cite{malakar2018steady},
\begin{align}
    \zeta\dot x_i
     &=
     -\lambda q_1\mathcal F'(x_i)
     +\zeta v_0\sigma_i(t),
     \qquad i=1,\ldots,N_b.
     \label{eq:eom-x}
\end{align}
The activity is characterized by the propulsion velocity $v_0$ and the telegrapher's noise $\sigma_i(t)=\pm1$ flips at rate $\alpha$. The autocorrelations of this random noise is exponential $\la\sigma(t)\sigma(0)\ra=\exp(-|t|/\tau)$, where $\tau=1/(2\alpha)$ is the persistence time. 

The interaction potential leads to a stochastic force $-\lambda\sum_i\mathcal F(x_i)$ by the active particles on the first oscillator, which in turn, exerts the force $-\lambda q_1\mathcal F'(x_i)$ on each of the active particles. We take $\mathcal F$ to be localized around one point of the ring, so that the interaction is short-ranged. A particle then acts on the chain only while it passes through the contact region, and each passage delivers a kick to the first oscillator.\\

At long times, the system reaches a nonequilibrium stationary state, which is characterized by a stationary energy current that flows through the chain. The energy current flowing through the chain is given by,
\begin{align}
    J=\langle \mathcal{J}(t)\rangle=\left\la \dot q_1\left(-\lambda\sum_{i=1}^{N_b}\mathcal F(x_i)   \right)\right\ra = -\lambda\,N_b\left\la \dot q_1\left(\mathcal F(x_i)   \right)\right\ra
\end{align}
where $\mathcal{J}(t)$ denotes the instantaneous current, and the averages $\la\cdot\ra$ are taken in the nonequilibrium stationary distribution of the joint dynamics~\eqref{eq:eoms}--\eqref{eq:eom-x}. The Hamiltonian nature of the bulk dynamics ensures that this equals the current flowing from  the $j$th to the $(j+1)$th oscillator, 
\begin{equation}
  J=\frac{k}{2m}\Big\la(p_j+p_{j+1})(q_j-q_{j+1})\Big\ra,
    \label{eq:current}
\end{equation}
We mainly focus on the case where the right reservoir is cold, $T_N=0$, so that the current is generated by the reservoir with the active particles.\\

We choose the contact kernel $\mathcal{F}$ to be of the von Mises form
\begin{equation}
    \mathcal F(x)
    =
    \frac{1}{I_0(p)}
    \exp\left[p\cos\left(\frac{2\pi x}{L}\right)\right],
    \label{eq:von-mises}
\end{equation}
where $I_0$ is the zeroth modified Bessel function. The parameter $p$ controls the localization of the interaction. The limit $p\to\infty$ approaches a periodic delta function, mimicking a point-like contact.

\section{Bath-induced forces on a single probe}\label{sec:probe}
Before studying the full chain, it is useful to isolate the effect of the active medium on a single slow probe coordinate $q(t)$, which is a representative of the left-most particle of the harmonic chain. We will work in the weak coupling limit where the coupling parameter $\lambda\ll 1$.
To that end we consider a probe $q(t)$ of mass $m$ in a harmonic trap of stiffness $K$, coupled to the $N_b$ particles in the same way as $q_1$ in Eq.~\eqref{eq:eom-q1},
\begin{align}
    m\ddot q
     &=
     -Kq+
     F_A(t) +\Lambda,\qquad F_A(t) = -\lambda\sum_{i=1}^{N_b}\mathcal F(x_i(t)).
    \label{eq:FA}
\end{align}
With the mobility $\mu\equiv\zeta^{-1}$, the particles obey $\dot x_i=-\mu\lambda q\,\mathcal F'(x_i)+v_0\sigma_i$. Assuming that the probe is slow compared with the relaxation of the medium, i.e.\
$\varepsilon= \omega_0t_{\mathrm b}\ll 1$ where $\omega_0=\sqrt{K/m}$ is the intrinsic frequency scale of the probe, and $t_{\mathrm b}=2\tau/\left(1-\mathrm{Re}\sqrt{1-\left(4\pi v_0\tau/L\right)^2}\right)$ is the relaxation time of the slowest
density mode of the active medium (see App.~\ref{app:validity} for a detailed discussion),
we rewrite this force as,
\begin{equation}
    F_A(t) = \la F_A(t)\,|\,q(s),\,s\le t\ra + \eta(t).
    \label{eq:FA-split}
\end{equation}
Here the first term denotes the average over the medium
dynamics with a prescribed probe history $q(s)$, $s\le t$.
Its response expansion gives the mean force associated with that
history. The fluctuating part is
$\eta(t)=F_A(t)-\la F_A(t)\,|\,q(s),\,s\le t\ra$.
To leading order in the weak coupling, this fluctuating force is
evaluated in the uncoupled stationary medium, where it has zero mean.
The mean force contains a quasistatic contribution and a friction term,
\begin{equation}
   \la F_A(t)\,|\,q(s),\,s\le t\ra
    = F^\qs(q(t)) + F^\fric(t).
    \label{eq:response-split}
\end{equation}
Here  $F^\qs (q) = \la F_A\ra_q$ is the mean force on a probe frozen at $q$, averaged over the stationary ensemble $\rho_q(x)$ at that position.  That is the quasistatic expression
\begin{align}
    F^\qs =-\lambda N_b\int_0^L\dd x\,\mathcal F(x)\,\rho_q(x)
\end{align}
To leading order in the weak coupling $\lambda$, this is computed in  App.~\ref{app1}:
\begin{align}
    F^\qs=-\lambda N_b+\kappa\,q,
    \qquad\text{with }
    \kappa = \frac{\lambda^2N_b}{\zeta v_0^2\tau}\left(\la\mathcal F^2\ra_0-\la\mathcal F\ra_0^2\right)
    = \frac{\mu N_b\lambda^2}{v_0^2\tau}\left[\frac{I_0(2p)}{I_0(p)^2}-1\right],
    \label{eq:kappa}
\end{align}
where $\la\cdots\ra_0$ is the average taken with respect to the uniform distribution on the ring. It is important to note that $\kappa>0$, implying that the medium weakens the trap. The constant part in the quasistatic force does not play a role in the energy conduction, so we choose $\Lambda=\lambda N_b$ in Eq.~\eqref{eq:eom-q1} to cancel it.

The remaining part in \eqref{eq:response-split}, $F^\fric$, arises due to the response in the medium due to a change in position of the probe,
\begin{align}
    F^\fric(t)&=\la F_A(t)\,|\,q(s),\,s\le t\ra- F^\qs \nonumber=\int_{-\infty}^t\dd s\,R(t-s)\left[q(s)-q(t)\right].
    \label{eq:fricdef}
\end{align}
where $R(t)$ is the response function measuring the change in the mean medium force at time $t$ caused by a displacement of the probe at an earlier time $s$. Integrating by parts, the above equation reduces to
\begin{equation}
    F^\fric(t) = -\int_{-\infty}^t\dd u\,\Gamma(t-u)\,\dot q(u),
    \qquad\text{with }
    \Gamma(u) = \int_u^\infty\dd s\,R(s).
    \label{eq:memory-kernel}
\end{equation}
To leading order, the response function can be obtained from the Kubo-Agarwal formula~\cite{agarwal1972fluctuation} as a correlation in the free medium,
$    R(t) = -\mu\lambda^2N_b\,\la\mathcal F(x(t))\,\mathcal F''(x(0))\ra_0$ [see~App.\ref{app1} for details].  Using it in \eqref{eq:memory-kernel}, the friction kernel $\Gamma$ is obtained in frequency space\footnote{Fourier transform is defined as $\tilde f(\omega)=\int dt \,e^{i\omega t} f(t)$},  
\begin{align}
\tilde\Gamma(\omega)=\mu\lambda^2\sum_{n\geq1}a_nk_n^2\,
\frac{1-\Omega_n^2\tau^2-i\,\omega\tau}{\Omega_n^2\left(\Omega_n^2\tau^2-\omega^2\tau^2-i\,\omega\tau\right)},
\end{align}
where $\Omega_n=k_nv_0$, with $k_n=2\pi n/L$, and $a_n=2N_bF_n^2$ is the weight of that harmonic, with $F_n= I_n(p)/I_0(p)$ the Fourier coefficients of $\mathcal F$.

It is interesting to note that, the real part of the friction kernel, which determines the dissipation,
\begin{equation}
\mathrm{Re}\,\tilde\Gamma(\omega)=\mu\lambda^2\sum_{n\geq1}a_nk_n^2\,\frac{\omega^2\tau^2+1-\Omega_n^2\tau^2}{(\Omega_n^2-\omega^2)^2\tau^2+\omega^2},
    \label{eq:ReGamma}
\end{equation}
is not always positive. In fact, for 
 $\omega^2<(\Omega_1^2\tau^2-1)/\tau^{2}$, all the terms in sum on the rhs are negative, implying that at these frequencies the induced friction, which normally dissipates energy, pumps energy into the probe.

 Also the time-integrated friction  coefficient $\gamma_\eff$, obtained by putting $\omega\to0$ \begin{align}
\gamma_\eff=\mu\lambda^2\sum_{n\geq1}a_n\,\frac{1-\Omega_n^2\tau^2}{k_n^2v_0^4\tau^2}
\end{align}
becomes negative in that regime. In this case, for a single probe, the linear
dynamics then has no stationary state, and nonlinear terms beyond order $\lambda^2$
are required to stabilize it.\\

The noise $\eta=F_A-\la F_A\,|\,q(s\le t)\ra$ already carries a factor $\lambda$. To leading order its covariance can therefore be evaluated in the free medium, 
\begin{align}
\la\eta(t)\eta(t')\ra=\lambda^2N_b\la\mathcal F(x(t));\mathcal F(x(t'))\ra_0.
\end{align}
In the frequency domain,
\begin{align}
\la    \tilde\eta(\omega) \tilde\eta(\omega')\ra=2\pi\delta(\omega+\omega')S(\omega)\quad \text{with }S(\omega)=\lambda^2\sum_{n\geq1}\frac{2a_n\,\Omega_n^2\tau}{(\Omega_n^2-\omega^2)^2\tau^2+\omega^2}.
    \label{eq:noisecorrS}
\end{align}
In the time-scale-separated regime, $\eta$ acts as a white noise with correlations $\la\eta(t)\eta(t')\ra\simeq S(\omega=0)\,\delta(t-t')$, where $    S(0)=\lambda^2\sum_{n\geq1}\frac{2\,a_n}{\Omega_n^2\tau}$. \\

Summarizing, the reduced dynamics of the probe is, to $O(\lambda^2)$,
\begin{equation}
    m\ddot q(t)
    =
    -(K-\kappa)\,q(t)
    -\int_0^\infty \dd s\,
    \Gamma(s)\dot q(t-s)
    +\eta(t)
    +O(\lambda^3),
    \label{eq:single-probe-gle}
\end{equation}
where the constant part of $F^\qs$ has been cancelled by $\Lambda$.
In the following sections, we study the energy transport problem of the full chain. We use this reduced description for the left most particle of the chain i.e., $q_1$ in Eq.~\eqref{eq:eom-q1} for analytical predictions and comparisons with simulations of the full system dynamics.

\section{Stationary currents}\label{sec:current}
The dynamics of the reduced chain, i.e., using \eqref{eq:single-probe-gle}, can be cast in a matrix form, 
\begin{align}
    m\ddot{\bm q}(t) = -\Phi\,\bm q(t)
    - \int_0^\infty\dd s\,\left[\Gamma(s)\,E_1+2\gamma_N\,\delta(s)\,E_N\right]\dot{\bm q}(t-s)
    + \bm\xi(t).
    \label{eq:chain-matrix}
\end{align}
Here $\bm q=(q_1,\dots,q_N)^T,$, $(E_i)_{j\ell}=\delta_{ij}\delta_{i\ell}$, $\Phi=\Phi_0-\kappa E_1$ where  $\phi_0$ is the force matrix of the bare chain without any reservoirs with matrix elements
\begin{align}
  \qquad
  (\Phi_0)_{j\ell}=\begin{cases}
    K+k, & j=\ell\in\{1,N\},\\
    2k,  & j=\ell\in\{2,\dots,N-1\},\\
    -k,  & |j-\ell|=1,\\
    0,   & \text{otherwise}.
  \end{cases}
\end{align}
$\bm\xi_i=\eta\delta_{i1}+\sqrt{2\gamma_NT_N}\,\eta_N\delta_{iN}$ denotes the noise vector. Note that, since we are interested in the active medium, for the remainder of the manuscript we will consider $T_N=0$. To compute the stationary current, we note that the formal solution to the reduced dynamics is best expressed in the frequency space
\begin{align}
    \tilde{\bm{q}}(\omega)=G(\omega)\bm{\xi}(\omega),\label{eq:greenf}
\end{align}
and $G(\omega)$ is the Green function,
\begin{align}
    G(\omega) = \left[\Phi_0 - m\omega^2 I
    - \left[\kappa+ i\omega\tilde\Gamma(\omega)\right]E_1-  i\omega\gamma_N\,E_N\right]^{-1}
\end{align}
with $I$ being the identity matrix. The stationary current corresponding to Eq.~\eqref{eq:chain-matrix} is given by the Landauer-B\"uttiker formula~\cite{dhar2001heat,DharRoy,santra2022activity} 
\begin{align}
     J = \frac{\gamma_N}{\pi}\int_0^\infty\dd\omega\,\omega^2\,|G_{1N}(\omega)|^2\,S(\omega), 
    \label{eq:J-negf}
\end{align}
 It is worth pointing out that in the previous studies~\cite{santra2022activity,sarkar2023stationary,sarkar2024rubin} activity entered only through the noise spectrum $S(\omega)$, while in the current scenario activity affects also the conduction spectrum. We evaluate the above form numerically and compare with numerical simulations of Eqs.~\eqref{eq:chain-all}-\eqref{eq:eom-x}. 
    
    \subsection{Instabilities}\label{sec:instab}
Consider the chain without the active medium $\lambda=0$. Then the chain evolves via the Green function $G_0(\omega)=[\Phi_0-m\omega^2I-i\omega\gamma_NE_N]^{-1}$. All the poles of the Green function lie in the lower half of the complex $\omega$ plane, and thus the bare chain is stable, meaning the linear system of equations reach a stationary state for any stationary driving. Now, as discussed above, the coupling to the active medium results in an additional negative stiffness and a friction kernel whose dissipative part $\mathrm{Re}\,\tilde\Gamma(\omega)$ can be negative. In the following, we analyse how these induced forces change the stability of the reduced linear system Eq.~\eqref{eq:chain-matrix}.

All elements of $G(\omega)$ share the poles given by the zeros $z_j$ of $\mathcal D(\omega)=\det G^{-1}(\omega)$. Each zero corresponds to a mode of the chain evolving as $e^{- i z_jt}$, which decays if $\mathrm{Im}\,z_j<0$. Since the kernel $\Gamma(t)$ is real, $\mathcal D(-\omega^*)=\mathcal D(\omega)^*$, so the zeros are either purely imaginary or come in pairs $(z_j,-z_j^*)$. As the parameters vary, the zeros move continuously, and the dynamics becomes unstable when a zero crosses the real axis into the upper half-plane. A mode then grows exponentially, the linear dynamics~\eqref{eq:chain-matrix} has no stationary state, and Eq.~\eqref{eq:J-negf} no longer describes a stationary current. The actual dynamics, however, does have a stationary state and is characterized by the nonlinear terms.
    

To locate the instabilities, we rewrite $\mathcal{D}(\omega)$ as
\begin{align}
    \mathcal D(\omega)=\det[ G_0^{-1}(\omega)]\left(1-\mathcal{L}(\omega)\right)\quad\text{with }
    \mathcal{L}(\omega)=(\kappa+i\omega\tilde\Gamma(\omega))\,[G_0(\omega)]_{11}
\end{align}
 where we have used the explicit form of $E_1$. The zeros of $\det G_0^{-1}$ are the stable poles of the bare chain, and thus an instability can therefore arise only from zeros of $1-\mathcal L(\omega)$ in the upper half-plane. Since the zeros move continuously with the parameters, the boundary of stability is reached when such a zero lies on the real axis, i.e., when
\begin{align}
    \mathcal L(\omega)=\left(\kappa+ i\omega\tilde\Gamma(\omega)\right)[G_0(\omega)]_{11}=1
    \label{eq:unstable}
\end{align}
for some real $\omega$.
The instability condition Eq.~\eqref{eq:unstable} can be divided into two categories
\begin{enumerate}
    \item A purely imaginary zero crosses the real axis at $z_j=0$. The physical reason for this instability is that the induced linear stiffness from the active medium becomes large and negative. This can also be seen by putting $\omega=0$ in Eq.~\eqref{eq:unstable}, which leads to 
\begin{align}
    \kappa\,(\Phi_0^{-1})_{11}>1 .\label{kappainst}
\end{align}
We will refer to this as a stationary instability, where the chain is pushed towards a new rest position
    \item A pair of zeroes $(z_j,-z_j^*)$ crosses the real axis at $\pm \text{Re}\, z_j$. The physical reason of this instability is due to the fact that the induced linear friction becomes high negative, enough to overcome the damping at the other end of the chain. We will refer to this as oscillatory instability, as the unstable mode grows while
oscillating at the frequency $\mathrm{Re}\,z_j$.
\end{enumerate}
An illustration of the instabilities are done after the discussion of stationary currents in Sec.~\ref{sec:tau} around Fig.~\ref{fig:jphasespace}.


\subsection{Current \textit{vs} propulsion speed}\label{sec:v0}
\begin{figure}
    \centering
    \includegraphics[width=\linewidth]{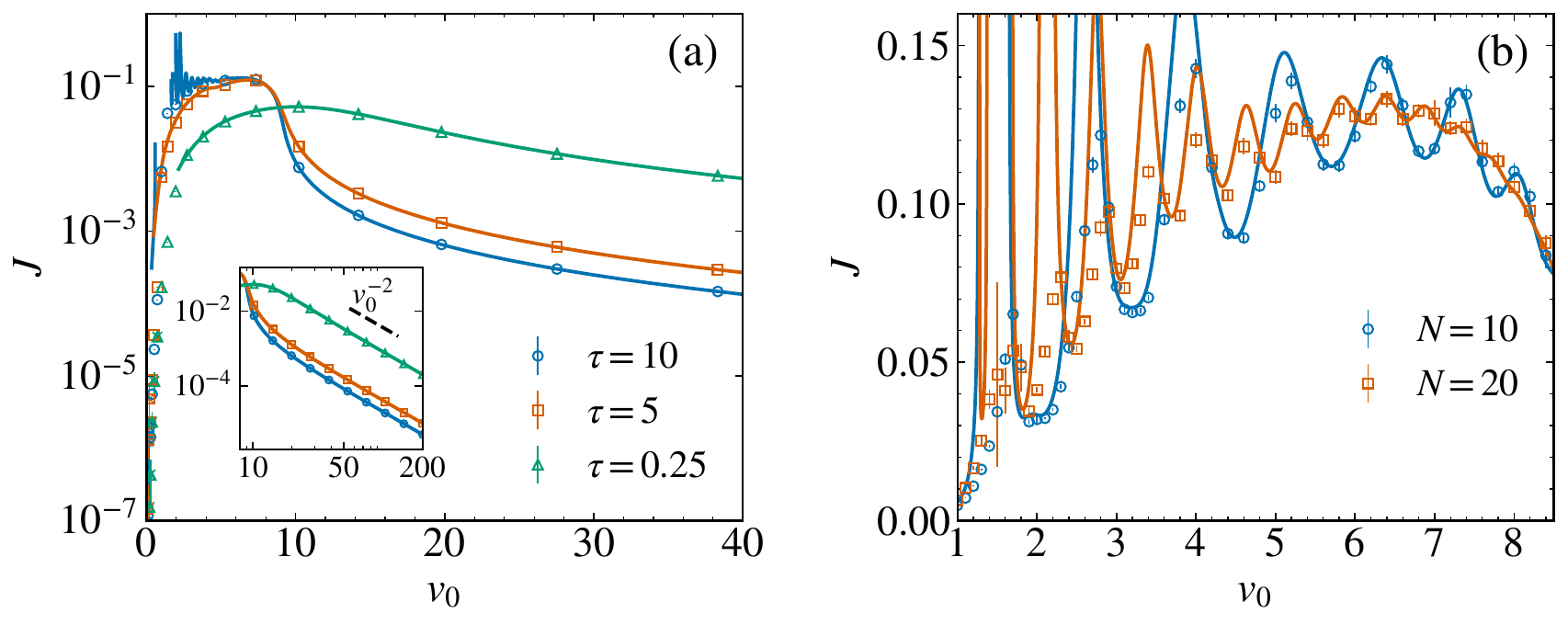}
    \caption{Average stationary current \textit{vs} propulsion speed. The main plot in (a) shows the variation of the current with $v_0$ for different values of persistence for a chain of length $N=32$. The inset shows the large-$v_0$ power law decay of the same data as in the main plot. Panel (b) shows the resonant behaviors near smaller $v_0$ and for $\tau=10$. In both panels the symbols denote the data obtained from numerical simulations of the full microscopic dynamics, while the solid lines denote the analytical prediction Eq.~\eqref{eq:J-negf} (integrals performed numerically). For both panels $m=k=K=\gamma_N=1,~\lambda=0.1,$ $N_b=10, \zeta=0.1, L=28,$ and $p=2$.}
    \label{fig:propvel}
\end{figure}

In this subsection, we discuss how the stationary current depends on the propulsion speed of the active particles of the medium. Figure~\ref{fig:propvel} (a) shows the stationary current as a function of the propulsion speed $v_0$ for different values of persistence $\tau=(2\alpha)^{-1}$. The current is very small for small $v_0$, increases to a plateau like region, followed by a two step decay.

We first focus on the increasing parts of the curve. We see that for highly persistent medium, oscillations appear, where the current is peaked at certain values of the propulsion velocity. To understand this, we note that the persistent medium dynamics has two characteristic frequencies, namely the frequency at which it crosses the ring $\Omega_1=2\pi v_0/L$, and the tumbling frequency $1/\tau=2\alpha$. So, for $v_0<v_c=L\omega_p/(2\pi)$ $\Omega_1$ lies within the phonon band $0<\omega<\omega_p=2\sqrt{k/m}$. 
In this region the peaks of $S(\omega)$ at $\Omega_1$, and possibly at its
harmonics $\Omega_n$, lie inside the phonon band [Fig.~\ref{fig:noisespectrum}(a)].
For a persistent medium, $\Omega_1\tau>1$, these peaks are sharp and the chain
is driven almost periodically. The current then shows resonances whenever
$\Omega_1$ equals a normal mode of the chain,
\begin{equation}
  v_0 \simeq \frac{L\omega_p}{2\pi}\,\sin\!\left(\frac{n\pi}{2(N+1)}\right),
  \qquad n=1,\dots,N.
\end{equation}
This is visible best for short chains ( see Fig.~\ref{fig:propvel} (b)), and the peaks
smooth out for larger $N$, as the spacing between consecutive normal modes goes
to zero. For less persistent media, $\Omega_1\tau<1$, the peaks of $S(\omega)$
merge into a smooth spectrum (green curve in Fig.~\ref{fig:noisespectrum}(a)), and
the resonances disappear.


For $v_0>v_c$, $\Omega_1$ lies above the phonon band. For a persistent medium,
$\Omega_1\tau\gg1$, the bath force is then close to a periodic drive at the
frequencies $\Omega_n\geq\Omega_1$, all outside the band. Its spectrum is
peaked at these frequencies, and tumbling gives the peaks a finite width
$\sim\tau^{-1}$. The chain therefore receives energy only through the tails of
these peaks that extend into the band. From Eq.~\eqref{eq:noisecorrS}, for
$\omega<\omega_p$,
\begin{equation}
  S(\omega)\simeq 2\lambda^2\sum_{n\geq1}\frac{a_n\,\Omega_n^2}{(\Omega_n^2-\omega^2)^2\,\tau},
\end{equation}
which decreases as $\tau^{-1}$. The current therefore drops sharply once
$v_0$ exceeds $v_c$, and the drop is steeper the larger $\tau$ is, see Fig.~\ref{fig:noisespectrum} (b).
This mirrors a harmonic chain driven periodically at one end, which responds
only when the driving frequency lies inside the phonon band, with resonances at
its normal modes~\cite{Prem2023KG}. For fast tumbling, $\Omega_1\tau\lesssim1$,
the spectrum is broad (see $\tau=0.25$ curve in Fig.~\ref{fig:noisespectrum} (b)) and no such drop occurs.
After the initial drop, the current has a slower power-law decay. These large
$v_0$ tails can also be estimated from Eq.~\eqref{eq:J-negf}. Taking
$v_0\to\infty$ at finite $\omega$, first in Eq.~\eqref{eq:noisecorrS}, yields
\begin{align}
    S(\omega)=\lambda^2\,2/(v_0^2\tau)\sum_n a_n/k_n^2+O(v_0^{-4}).
    \label{v0Lvc:1}
\end{align}
 The same limit for the Green function  yields,
 \begin{align}
     G^{-1}\to [\Phi_0-m\omega^2I- i\omega\gamma_NE_N+ O(v_0^{-2}) ].\label{v0Lvc:2}
 \end{align} 
 Combining these in Eq.~\eqref{eq:J-negf}, we see that the current is predicted to decay as $v_0^{-2}$. This behavior is verified in the inset of Fig.~\ref{fig:propvel} (a). It is interesting to note that, Eqs.~\eqref{v0Lvc:1} and~\eqref{v0Lvc:2} show that in this limit the current depends on the activity only through the well-known effective temperature of RTPs, , $T_{\rm RT}=\zeta v_0^2\tau$,\footnote{Equivalently,
$T_{\rm RT}=\zeta D_{\rm RT}$, with $D_{\rm RT}=v_0^2\tau$ the long-time diffusivity of a free
run-and-tumble particle.} namely $J\propto T_{\rm RT}^{-1}$. This behavior is analogous to the high-temperature
negative differential conductivity found for a passive contact
bath in \cite{ndc2026}. 

\subsection{Current \textit{vs} persistence}\label{sec:tau}
\begin{figure}
    \centering
    \includegraphics[width=\linewidth]{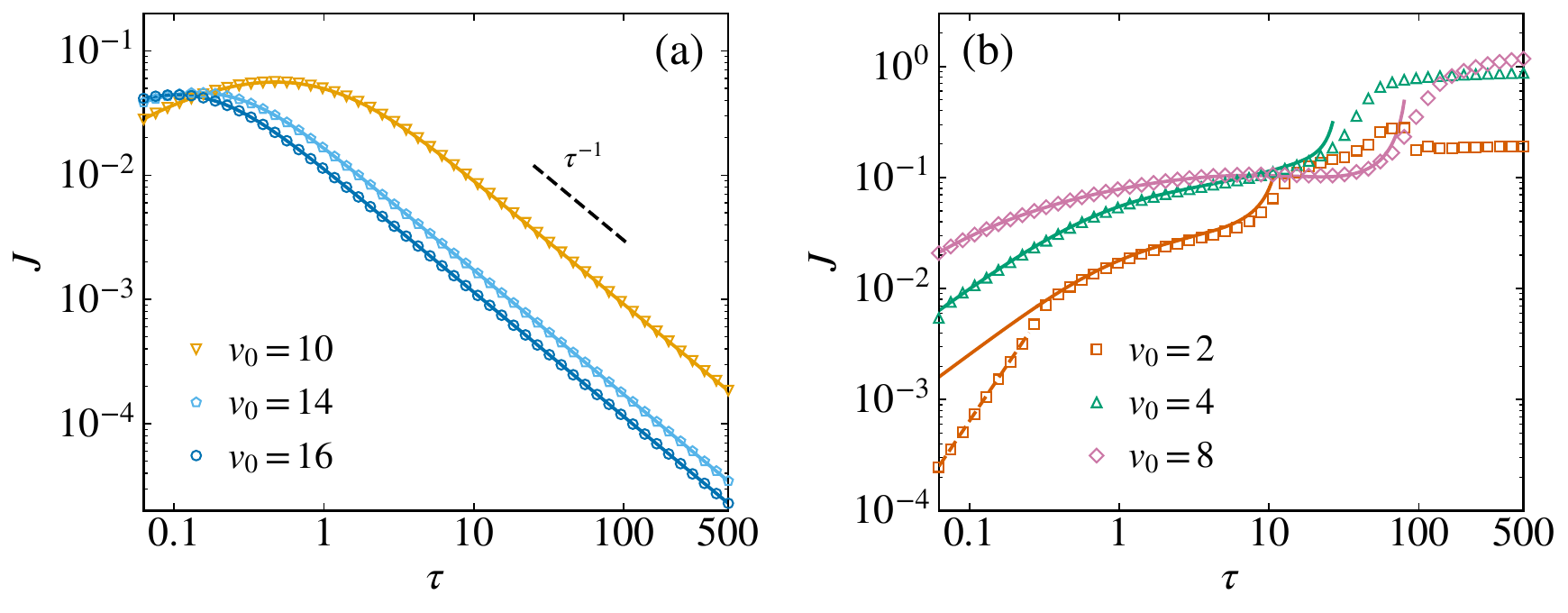}
    \caption{Average stationary current \textit{vs} persistence time for (a) $v_0>v_c\approx 8.9$ and (b) $v_0<v_c$. The symbols denote the data obtained from numerical simulations, while the solid lines denote the numerical integration of \eqref{eq:J-negf}. In panel (b) the dashed line (at small $\tau$) for $v_0=2$ correspond to the prediction \eqref{eq:unstableJAPS}. For both plots $N=32$, $m=k=K=\gamma_N=1,~\lambda=0.1,$ $N_b=10, \zeta=0.1, L=28,$ and $p=2$.}
    \label{fig:jvsactivity}
\end{figure}

In this section, we discuss how the average stationary current changes with the persistence time $\tau$. Figure~\ref{fig:jvsactivity} shows current as a function of activity for different  values of the propulsion speed $v_0$. We see that there are two distinctly different behaviors, depending on if $\Omega_1$ is greater or lesser than $\omega_p$, or in other words between the competition of $v_0$ and $v_c.$

\begin{figure}
    \centering
    \includegraphics[width=0.85\linewidth]{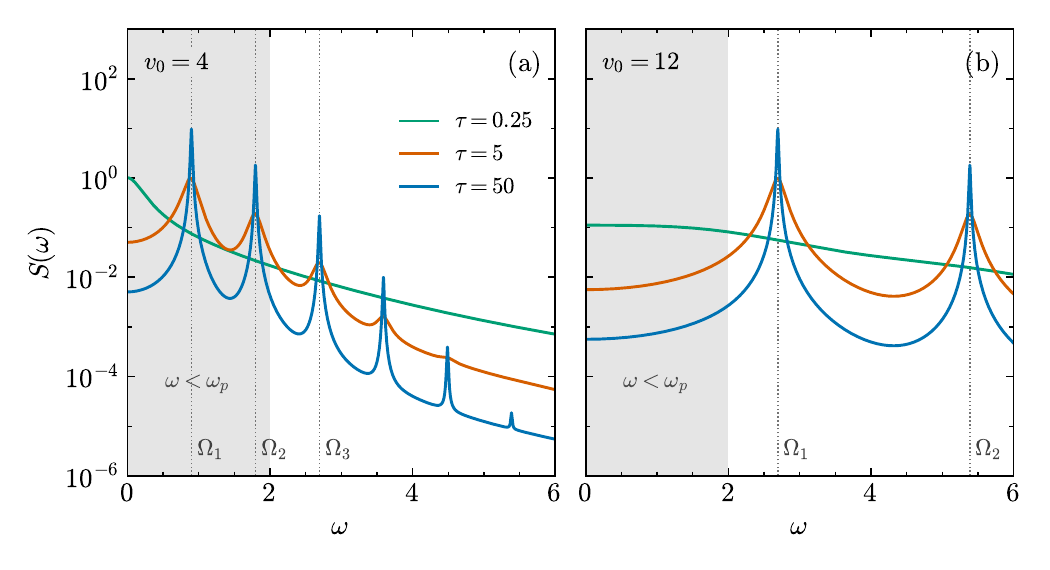}
    \caption{Bath spectrum $S(\omega)$ from Eq.~\eqref{eq:noisecorrS}, for
(a) $v_0=4<v_c$ and (b) $v_0=12>v_c$ at three different persistence time-scales for the same parameters as in Fig.~\ref{fig:propvel}. The shaded region is the phonon band $\omega<\omega_p$ and the dotted lines mark the crossing
frequencies $\Omega_n=2\pi n v_0/L$. For $v_0>v_c$ the chain receives only the
tails of the peaks, which decrease as $\tau^{-1}$. }
    \label{fig:noisespectrum}
\end{figure}

For $v_0>v_c$, the characteristic frequency $\Omega_1$ lies outside the phonon band. In this limit, the bath spectrum follows Eq.~\eqref{v0Lvc:1}. Thus, increasing persistence reduces the spectral weight of the bath at the frequencies transmitted by the chain. Moreover, since the Green function Eq.~\eqref{v0Lvc:2} approaches a $\tau$ independent limit, the current decreases. The decay can be obtained by using Eqs.~\eqref{v0Lvc:1} and \eqref{v0Lvc:2}
 in Eq.~\eqref{eq:J-negf}, we see that the current for large $\tau$ decays as $\tau^{-1}$.

On the other hand, for $v_0<v_c$, the current increases with $\tau$ and
saturates for very large persistence. Here the peaks of $S(\omega)$ lie inside
the phonon band, and their height, $S(\Omega_n)=2\lambda^2 a_n\tau$, grows with
persistence (see Fig.~\ref{fig:noisespectrum}(a)). For large $\tau$, $\mathrm{Re}\,\tilde\Gamma(\omega)$, Eq.~\eqref{eq:ReGamma}, is negative below the crossing frequencies, and the linear theory becomes unstable once this negative friction overcomes the damping of the chain (oscillatory instability). This is also seen in Fig.~\ref{fig:jvsactivity} (b), where the theoretical curves corresponding to Eq.~\eqref{eq:J-negf} start deviating from the simulations for large $\tau$. However, as $\tau\to\infty$, the medium particles move ballistically, and the tumbling time exceeds all other time scales, so the current becomes independent of the persistence. The saturation value increases with $v_0$ up to $v_c$. The largest currents are thus obtained for highly persistent media with $\Omega_1$ inside the phonon band. 

The current also goes to zero at small persistence. However, in this regime too, the linear theory fails, as seen for $v_0=2$ in Fig.~\ref{fig:jvsactivity} (b). 
This can be understood in the following way. The induced negative stiffness $\kappa\propto(v_0^2\tau)^{-1}$, diverges in this limit. Once $\kappa(\Phi_0^{-1})_{11}>1$, the linearized dynamics around the origin
is unstable, \emph{stationary instability}. This motivates an expansion about a localized state with a nonzero displacement $q_1^*$.

Further, we note that when $\tau\to 0,$ the tumbling happens frequently and the bath noise emulates a white noise with strength $\sqrt{2D_{\mathrm{RT}}}=\sqrt{2v_0^2\tau}$. The bath particles then follow,  
\begin{align}
    \dot x_i=-\mu\lambda q_1\mathcal F'(x_i)+\sqrt{2D_{\mathrm{RT}}}\,\xi_i(t),
    \qquad\langle\xi_i(t)\xi_j(t')\rangle=\delta_{ij}\delta(t-t').
    \label{app:smalltau-bath}
\end{align}
In this $D_{\mathrm{RT}}\to0$ limit, we look for a rest configuration $\boldsymbol q^*=(q_1^*,\ldots,q_N^*)^{T}$, with the medium particles localized at $x^*$, where $\mathcal{F}'(x^*)=0$. The rest position $\boldsymbol q^*$ follows from force balance on the chain, i.e., putting $\ddot q_i=\dot q_i=0$ in Eqs.~\eqref{eq:eoms}. This yields $\left[ \Phi_0\boldsymbol q^*\right]_i
       =\lambda N_b[1-\mathcal F(x^*)]\delta_{i1}$, i.e.,
\begin{align}
     \quad q_1^*=\lambda N_b  [\Phi_0^{-1}]_{11}[1-\mathcal F(x^*)],
\end{align}
For the von Mises profile with $p>0$, there are two locally stable
possibilities,
\begin{align}
    x^*=L/2:\qquad
    q_{1,+}^*&=\lambda N_b  [\Phi_0^{-1}]_{11}
          \left[1-\frac{e^{-p}}{I_0(p)}\right]>0,
          \label{app:smalltau-positive}\\
    x^*=0:\qquad
    q_{1,-}^*&=\lambda N_b  [\Phi_0^{-1}]_{11}
          \left[1-\frac{e^{p}}{I_0(p)}\right]<0.
          \label{app:smalltau-negative}
\end{align}
The second derivative $\mathcal{F}''(x)=k_1^2\mathcal{F}(x)(p^2\sin^2(k_1x)-p\cos(k_1x))$ positive for $x^*=L/2$, and negative for $x^*=0$, implying $\mu\lambda q_1^*\mathcal F''(x^*)>0$. For the parameters used here,
$q_{1,-}^*\mathcal F''(0)>
q_{1,+}^*\mathcal F''(L/2)$,
so the bath particles are more tightly confined on the negative
branch. We therefore use $q_1^*=q_{1,-}^*$ and $x^*=0$
for the small-persistence estimate below.
Writing $x_i=x^*+y_i$ and expanding the bath dynamics to first
order in $y_i$, 
\begin{align}
     \dot y_i=-\mu\lambda q_1^*\mathcal F''(x^*)\,y_i+\sqrt{2D_{\mathrm{RT}}}\,\xi_i(t),
    \qquad
    \langle y_i(t)y_j(0)\rangle
      =\delta_{ij}\frac{D_{\mathrm{RT}}}{\mu\lambda q_1^*\mathcal F''(x^*)}e^{-\mu\lambda q_1^*\mathcal F''(x^*)|t|}.
\end{align}
Expanding the contact force to second order in $y_i$,
Eq.~\eqref{eq:eom-q1} becomes
\begin{align}
    m\ddot q_1
     &=
     -Kq_1-k(q_1-q_2)
     -\left[\lambda\sum_{i=1}^{N_b}\left(\mathcal\ F(x_i^*)+\mathcal{F}''(x_i^*)\frac{\la y_i^2\ra}{2}\right)-\Lambda\right]-\frac{\lambda}{2} \sum_{i=1}^{N_b} \mathcal{F}''(x_i^*)(y_i^2-\la y_i^2\ra).\label{eqapp:q1}
\end{align}
The term within the brackets give a constant force, and does not affect the current, while the last term $\Sigma=-(\lambda/2) \sum_{i=1}^{N_b} \mathcal{F}''(x_i^*)(y_i^2-\la y_i^2\ra)$ acts as a noise  with stationary correlations,
\begin{align}
   C_\Sigma(t)= \la\Sigma(t)\Sigma(0)\ra=\frac{N_b\lambda^2\mathcal F''(x^*)^2}{2}\,\frac{D_{RT}^2}{[\zeta^{-1}\lambda q^*\mathcal F''(x^*)]^2}\,e^{-2[\zeta^{-1}\lambda q^*\mathcal F''(x^*)]|t|}\label{eqqpp:q12}
\end{align}
So we have a new set of linear equations in this regime, with the equation of motion of $q_1$ satisfying Eqs.~\eqref{eqapp:q1} and \eqref{eqqpp:q12}. The stationary current still satisfies the general form in Eq.~\eqref{eq:J-negf},  
\begin{align}
    J = \frac{\gamma_N}{\pi}\int_0^\infty\dd\omega\,\omega^2\,|G'_{1N}(\omega)|^2\,S'(\omega),
    \label{eq:unstableJAPS}
\end{align}
with a Green function $G'=[\Phi_0-m\omega^2I-i\omega\gamma_NE_N]^{-1}$, and $S'(\omega)=\tilde C_\Sigma(\omega)$. Thus the observed current is proportional to $D_{RT}^2=v_0^4\tau^2$. The modified formula for current can be evaluated numerically in the unstable region and shows very good agreement.


\begin{figure}
    \centering
    \includegraphics[width=\linewidth]{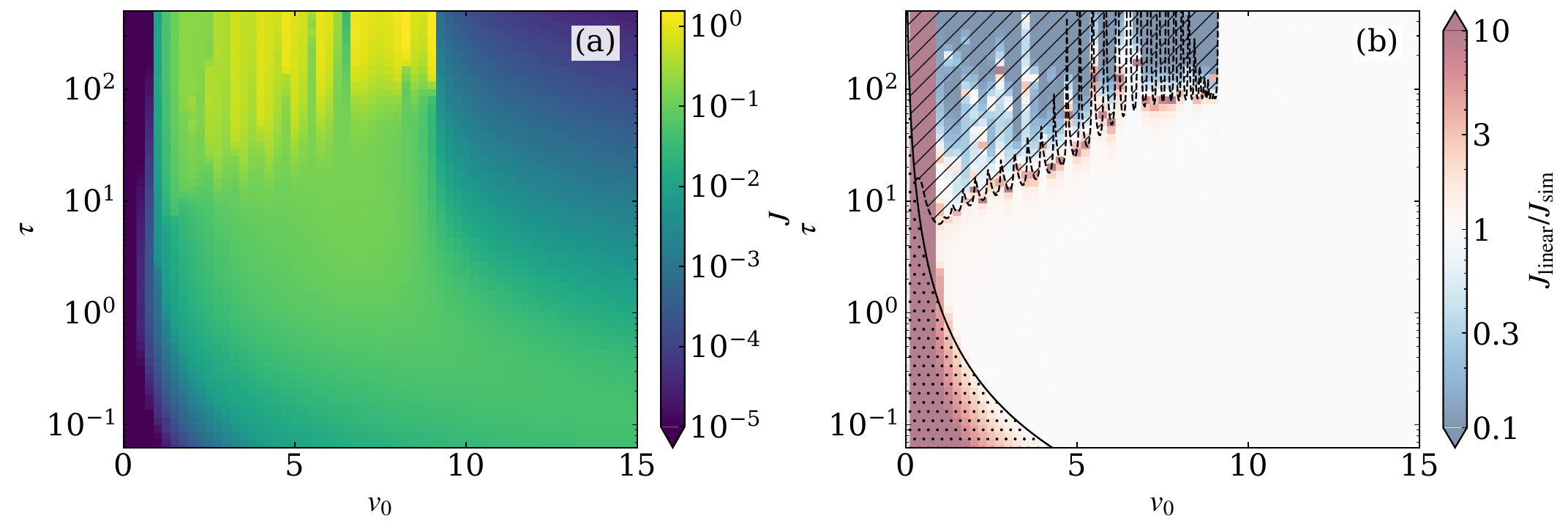}
    \caption{Panel (a) shows the average stationary current in the $(v_0,\tau)$ plane as obtained from numerical simulations. Panel (b) shows a ratio of the same data with that obtained from Eq.~\eqref{eq:J-negf}. The solid and dashed lines corresponds to  Eqs.\eqref{kappainst}, and  \eqref{eq:unstable} (for $\omega\neq 0$). For both panels, we have taken $N=32$, $m=k=K=\gamma_N=1,~\lambda=0.1,$ $N_b=10, \zeta=0.1, L=28,$ and $p=2$.}
    \label{fig:jphasespace}
\end{figure}

To end the discussion on the stationary current, we present a contour plot of the current in the $(\tau,v_0)$ plane in Fig.~\ref{fig:jphasespace} (a). The current is largest for highly persistent media where propulsion velocity $v_0$ is within that set by the characteristic frequency of the phonon modes $v_c$. In this limit of large persistence, the current shows a sharp drop beyond $v_0>v_c$. The linear theory fails in two regions consistent with the general predictions from Eq.~\eqref{eq:unstable}. The zero frequency crossing occurs when the linear induced stiffness $\kappa\propto (v_0^2\tau)^{-1}$ becomes very large. The condition Eq.\eqref{kappainst} marks this region shaded with dots. The other instability occurs due to induced negative linear friction, the boundary in this case Eq.~\eqref{eq:unstable} for $\omega\neq 0$ is marked with dashed lines, and the linearly unstable region is shaded with striped solid lines. In between these two boundaries the effective linear theory from reduced dynamics captures the stationary current very well.

\section{Bulk properties of the conducting chain}\label{sec:bulk}
We now want to understand how the presence of the active medium reflects on the stationary bulk properties of the conductor. In particular, we look at the kinetic temperature profile, the single-oscillator velocity distributions, and the spatial velocity correlations.

\subsection{Temperature profile}
\begin{figure}
    \centering
    \includegraphics[width=0.75\linewidth]{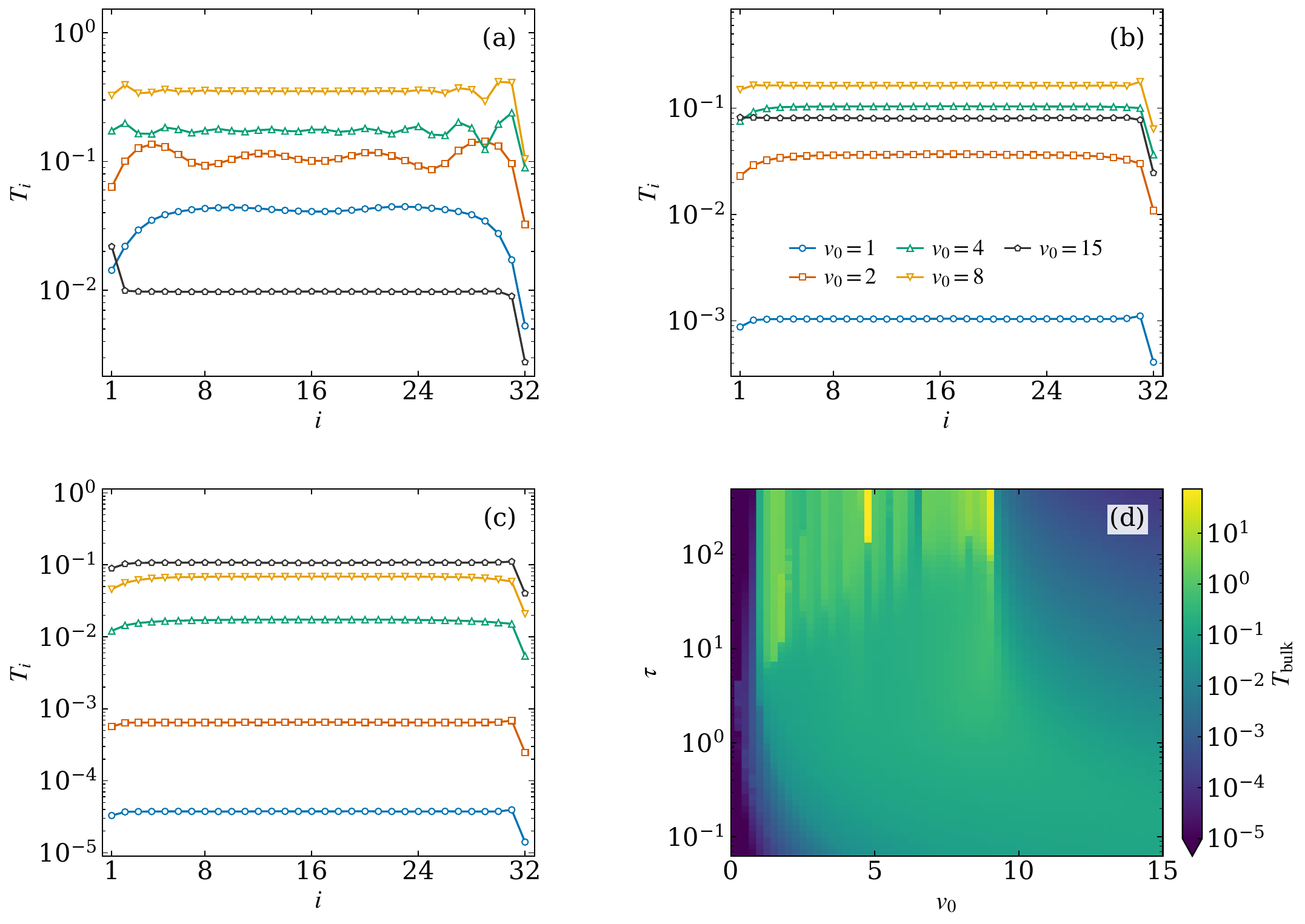}
    \caption{Kinetic temperature profile in the conductor. Panel (a), (b) and (c)  show the full temperature profile for $\tau=5,\,0.5,\,0.0625$ respectively. Panel (d) shows bulk temperature $T_\text{bulk}$ with $l_0=4$. For all the plots the data is obtained from numerical simulations with $N=32$, $m=k=K=\gamma_N=1,~\lambda=0.1,$ $N_b=10, \zeta=0.1, L=28,$ and $p=2$.}
    \label{fig:tempprof}
\end{figure}

We first look at the kinetic temperature profile of the harmonic chain,
\begin{align}
     T_i=m\la\dot q_i^2\ra,\qquad i=1,\ldots,N.
\end{align}
The bulk temperature profile is flat, with boundary layers at the two ends as shown in Fig.~\ref{fig:tempprof} (a)-(c). This is a consequence of the harmonic nature of the conductor which exhibits ballistic transport. The bulk temperature profile $T_\text{bulk}=\frac{1}{N-l_0}\sum_{i=l_0/2}^{N-l_0/2} T_i$ (where $l_0\ll N$ is chosen to avoid the boundary sites). Figure~\ref{fig:tempprof} (d) shows $T_\text{bulk}$ in the $(v_0,\tau)$ plane. The qualitative behavior shows the same trends as the current. The largest bulk temperature occurs for large $\tau$ and $v_0\lesssim v_c$, upon increasing $v_0$ beyond $v_c$, the current drops sharply. For small $\tau$, the effects are much less drastic, but also shows the same nature, namely, an initial increase with $v_0$, followed by a slow decrease at large values of $v_0$.

\subsection{Velocity distributions}
\begin{figure}
    \centering
    \includegraphics[width=\linewidth]{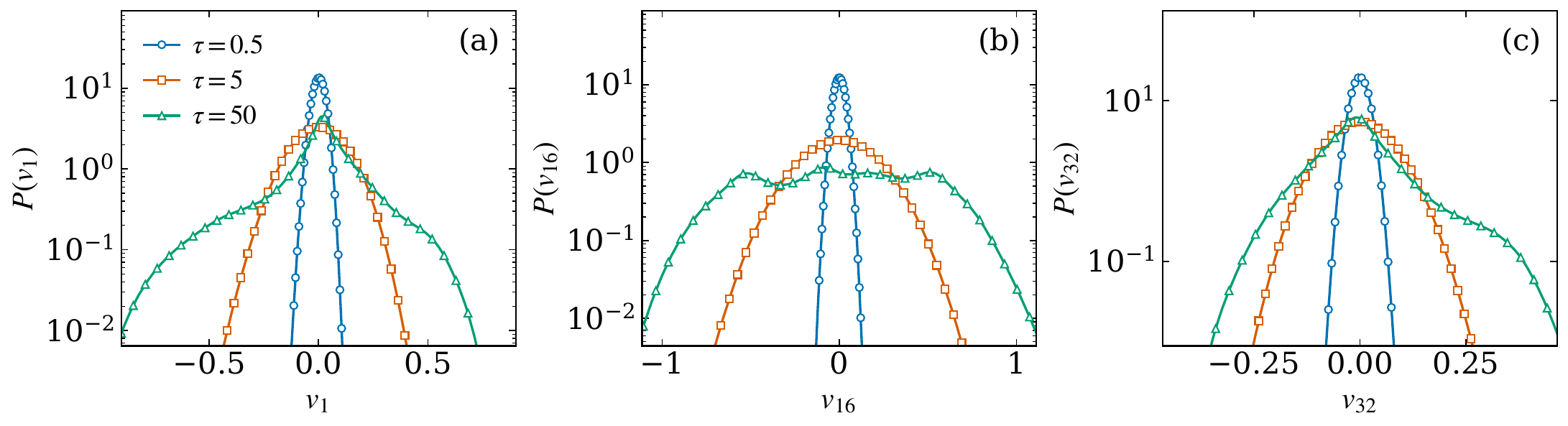}
    \caption{Velocity distributions $P(v_i)$ for $i=1,\,16,\,32$ for a chain of length $N=32$. The plots are obtained from numerical simulations of Eq.~\eqref{eq:chain-all}-\eqref{eq:eom-x} with $v_0=1$, $m=k=K=\gamma_N=1,~\lambda=0.1,$ $N_b=10, \zeta=0.1, L=28,$ and $p=2$.}
    \label{fig:veldist}
\end{figure}

The kinetic temperature profile gives a measure of the velocity fluctuations along the conductor. To understand the higher order fluctuations, we look at the velocity fluctuations in the bulk. In particular, we numerically look at the velocity distribution of the first, middle, and $N$th oscillator for $v_0<v_c$, these are shown in Fig.~\ref{fig:veldist}. We find that for high persistence, all the distributions are highly non-Gaussian. More interestingly, $P(v_1)$ is positively skewed with a single peak, $P(v_{N/2})$ shows a symmetric multi-peaked structure, while $P(v_N)$ shows a singly peaked negatively skewed distribution. It is worth noting that $\tau=50$ lies in the regime of oscillatory instability discussed in Sec.~\ref{sec:instab}, and indicates that nonlinearities are essential to capture these effects. For smaller values of $\tau$, the distributions can be well approximated by a Gaussian. This holds for both the stable region and the stationary instability.

\subsection{Correlations in the bulk}
The velocity distributions clearly indicate that the active medium generates non-trivial correlations along the conductor, especially in the region of instability. To understand this better we study the equal time spatial correlations of the oscillator velocities,
\begin{align}
      C_{j\ell}=\la\dot q_j\dot q_{\ell}\ra-\la\dot q_j\ra\la\dot q_\ell\ra\label{eq:corrdef}
\end{align}
For heat conduction between equilibrium Langevin reservoirs, the velocities of bulk oscillators are uncorrelated, $C_{j\ell }\propto\delta_{j\ell}.$ It was observed in previous studies of active energy conduction that oscillator media generates correlations in the conducting chain which decay exponentially~\cite{sarkar2024rubin}. Here we observe a very different structure.

\begin{figure}
    \centering
    \includegraphics[width=\linewidth]{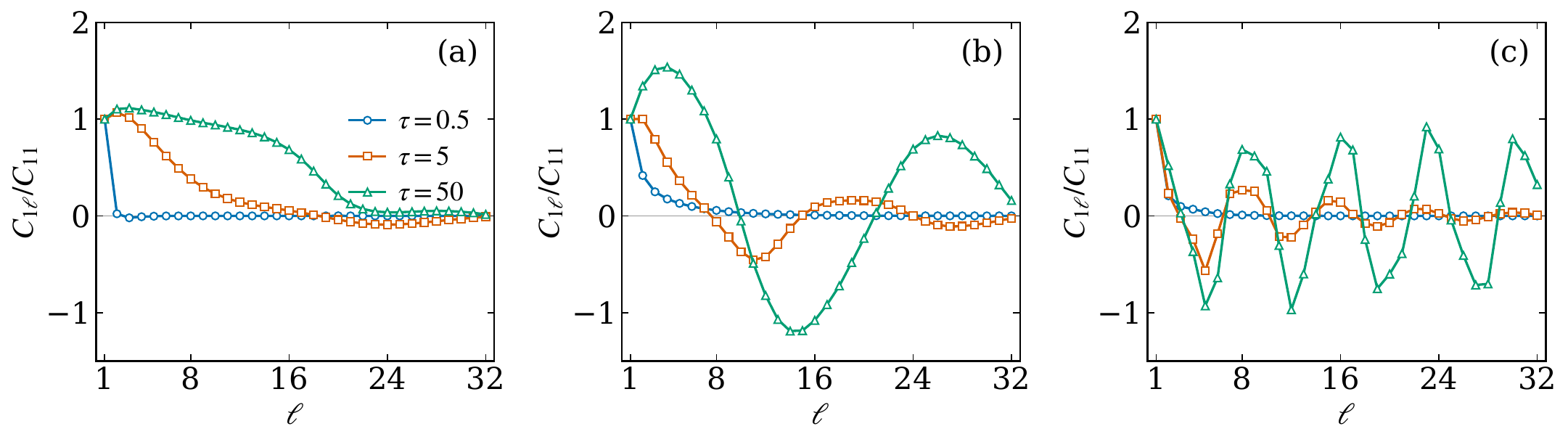}
    \caption{Connected velocity correlations along the chain, as defined in Eq.~\eqref{eq:corrdef}, keeping $j=1$, for a chain of length $N=32$. The plots are obtained from numerical simulations of Eq.~\eqref{eq:chain-all}-\eqref{eq:eom-x} with  $m=k=K=\gamma_N=1,~\lambda=0.1,$ $N_b=10, \zeta=0.1, L=28,$ $p=2$ for all panels; and $v_0=1,2,\,4$ for panels (a), (b), and (c), respectively.  }
    \label{fig:velcorr}
\end{figure}

Figure~\ref{fig:velcorr} shows $C_{1\ell}/C_{11}$ for three propulsion speeds, $v_0=1$, $2$ and $4$, all below $v_c$, and three persistence times. For small persistence $\tau=0.5$, the correlations decay to zero very fast for all chosen $v_0$. This holds both for the regions which are linearly stable (Figs.~\ref{fig:velcorr} (b) and (c)), and stationary instability ($v_0=1$, Fig.~\ref{fig:velcorr} (a)). For intermediate persistence $\tau=5$, the linear theory is stable for all three speeds. The correlations then extend further into the chain and, for $v_0=2$ and $4$, oscillate while decaying, similar to the behavior found for active Rubin media~\cite{sarkar2024rubin}. The oscillations are set by the chain mode that resonates with the crossing frequency $\Omega_1$, and become faster as $v_0$ increases.

For high persistence ($\tau=50$), all three speeds lie in the region of the oscillatory instability, and non-trivial correlations are generated along the length of the chain. For $v_0=1$ (Fig.~\ref{fig:velcorr} (a)), the velocities remain strongly and positively correlated with the active end for a long distance. For $v_0=2$ (Fig.~\ref{fig:velcorr} (b)), the correlations show oscillations, however damped, along the length of the chain. For $v_0=4$ (Fig.~\ref{fig:velcorr} (c)), the correlations exhibit undamped oscillations along the length of the chain. 

\section{Conclusion}\label{sec:concl}

We have studied energy transport through a harmonic chain driven at one end by a medium of run-and-tumble particles and drained at the other end by a zero-temperature Langevin reservoir. Inspired by soft-matter environments, the medium interacts with the chain through a short-range contact interaction. Our aim was to understand how the activity of the medium, characterized by the propulsion speed $v_0$ and the persistence time $\tau$, determines the energy flowing through the chain. 

To this end, we first find a reduced dynamics of the particle coupled to the active medium by integrating out the medium at weak system-medium coupling, with a time-scale separation as a sufficient condition; keeping the full memory of the medium, the result applies well beyond it. We compute the medium induced forces to second order in the weak coupling parameter and find that the induced stiffness $-\kappa$ is negative, with $\kappa$ proportional to $(v_0^2\tau)^{-1}$, the dissipative part of the memory friction becomes negative at frequencies below the ring-crossing frequencies once $2\alpha<\Omega_n$, the friction of a persistent medium then pumps energy into the chain instead of absorbing it.
The corresponding Landauer-B\"uttiker formula gives the stationary current quantitatively over a large part of the $(v_0,\tau)$ plane.
Its main merit, however, is that it delimits its own domain of validity.
The zeros of $\det G^{-1}(\omega)$ cross the real axis either at $\omega=0$, when the induced softening overcomes the chain (stationary instability, small $v_0^2\tau$), or at a pair $\pm\omega\neq0$, when the induced friction is negative enough to even overcome the damping from the other bath (oscillatory instability, large $\tau$ with $v_0<v_c$).
Both boundaries agree with where the linear prediction departs from the simulations.
In the first case the failure is mild: the chain relaxes to a shifted rest position, the medium particles become Ornstein--Uhlenbeck-like around the extrema of the coupling, and a linear theory around that state again predicts the current, now scaling as $D_\text{RT}^2$.
In the second case the linearly unstable modes are stabilized only by the nonlinearity of the coupling, and a reduced description must keep that nonlinearity.\\

The phenomenology of the current is organized by the position of the crossing frequency $\Omega_1=2\pi v_0/L$ relative to the phonon band.
For persistent media and $v_0<v_c$, the current is resonant with the normal modes of the chain and the largest currents are reached; for $v_0>v_c$ it drops sharply.
For fast tumbling, the dependence on $v_0$ is smooth.
At large propulsion speed and fixed persistence, the current
decays as $v_0^{-2}$, while at fixed $v_0>v_c$ it decays as
$\tau^{-1}$ for large persistence. Both decays follow from the
decreasing spectral weight of the bath force at the chain
frequencies and give rise to negative differential conductance.
Below $v_c$, the crossing frequency lies within the phonon band, and 
for highly persistent media, resonant forcing and negative
induced friction then produce the strongest transport.\\

The same regime leaves the clearest mark on the chain.
While the kinetic temperature profile stays flat in the bulk and follows the current, the velocity statistics in the region of the second instability are strongly non-Gaussian, and the equal-time velocity correlations extend over the whole chain, with damped or undamped spatial oscillations selected by the resonating chain mode.
None of this is visible for equilibrium reservoirs, and it goes beyond the exponentially decaying correlations generated by active oscillator mediums.\\

Several questions remain open.
A reduced theory for the oscillatory instability needs to go beyond linear order in the probe displacement, and should explain the saturation of the current and the non-Gaussian velocity statistics.
It would also be interesting to see which features survive for other active dynamics (active Brownian or active Ornstein--Uhlenbeck particles)~\cite{Santra2022universal}, for anharmonic conductors where transport is no longer ballistic~\cite{LepriLiviPoliti1997,Mai2007FPU,Lepri2009stochastic}, for a second active medium at the right end, and at the level of current fluctuations.\\
More generally, the example shows that an active environment acts on a system through all of its induced forces (stiffness, friction and noise), and that it is the interplay of these, not an effective temperature, that decides how much energy gets through.\\

\vspace{2cm}
\noindent {\bf Acknowledgment:}
IS thanks Bijay K Agarwalla for useful discussions on energy conduction in periodically driven chains. IS acknowledges funding from European Union’s
Horizon Europe research and innovation programme under the Marie Skłodowska–Curie (HORIZON-TMA-MSCA-PF-EF) grant agreement No. 101205210.

\appendix

\section{Validity of the decomposition of forces}\label{app:validity}
To understand the reduced dynamics, we decompose the bath forces in Eq.~\eqref{eq:FA-split}, as a quasistatic part, and a linear response about this state. In this section we discuss when such a decomposition should work. To this end, we first write the backward generator for the full dynamics,
\begin{align}
    \mathcal{L}&=\mathcal{L}^B+\mathcal{L}^P\\
    \text{with }\mathcal{L}^B
    &=\sum_{i=1}^{N_b}\left[
    \left(v_0\sigma_i-\mu\lambda q\mathcal F'(x_i)\right)\partial_{x_i}
    +(2\tau)^{-1}(T_i-1)\right]\\
    \mathcal{L}^P&=v\partial_q- m^{-1}\left(Kq+\lambda \left[\sum_{i=1}^{N_b}\mathcal F(x_i)-N_b\right]\right)\partial_v,
\end{align}
where the operator $T_i$ reverses the $i$th particle. 

Let $q_c>0$ be a characteristic displacement amplitude. The uncoupled
probe obeys $\ddot q=-\omega_0^2q$, where $\omega_0=\sqrt{K/m}$.
An oscillation of amplitude $q_c$ therefore has velocity amplitude
$v_s=\omega_0q_c$, corresponding to the momentum scale $mv_s=\sqrt{mK}\,q_c$. Let $t_{\mathrm b}$ denote the
relaxation time of the bath, determined below. In terms of the scaled variables
$y=q/q_c$, $w=v/v_s$, and $\bar t=t/t_{\mathrm b}$,
\begin{align}
     \widehat{\mathcal L}
 &=\widehat{\mathcal L}^{B}
   +\varepsilon\,\widehat{\mathcal L}^{P},
\end{align}
where 
\begin{align}
    \widehat{\mathcal L}^{B}=t_{\mathrm b}\mathcal L^{B},\qquad
    \widehat{\mathcal L}^{P}
 &=w\partial_y-\left[y+\frac{\lambda}{Kq_c}\left(\sum_{i=1}^{N_b}\mathcal F(x_i)-N_b\right)\right]\partial_w,~~\text{and }\qquad \varepsilon=t_{\mathrm b}\omega_0.
\end{align}

For the quasistatic response, we assume that the medium evolves with the probe frozen at a position $q$, with the stationary distribution $\rho_q(x)$. Here $\rho_q(x)=\sum_{\sigma=\pm1}P_q(x,\sigma)$, where $P_q$ is the stationary
solution of the medium dynamics at fixed $q$, $[\mathcal{L}^B]^\dagger P_q=0$. The mean bath force in this frozen stationary state defines the quasistatic contribution. 
However, the probe is not actually at rest, but has a trajectory $q(s)$ for $s\leq t$. The first correction to the quasistatic force therefore comes from the
difference between the actual trajectory and the frozen position, $q(s)-q(t)$.
We treat this difference as a perturbation and expand to first order in it. The perturbed bath generator is thus $\mathcal L^{B}+\mathcal L'^{B}$ with,
\begin{align}
    \mathcal L'^{B}
 =-\mu\lambda\,[q(s)-q(t)]\sum_i\mathcal F'(x_i)\partial_{x_i},\label{perturbapp}
\end{align}
which is proportional to the weak coupling parameter $\lambda$. To estimate $q(s)-q(t)=\int_t^s \mathrm{d}u\,v(u)$, we use the Cauchy--Schwarz
inequality and stationarity,
\begin{align}
    \left\langle[q(s)-q(t)]^2\right\rangle^{1/2}
 \leq (t-s)\,\langle \dot q^2\rangle^{1/2}
 =\varepsilon\,\frac{t-s}{t_{\mathrm b}}\,q_c
       \left\langle\left(\frac{v}{v_s}\right)^2\right\rangle^{1/2}.
\end{align}
Since the medium relaxes over a time $t_{\mathrm b}$ (determined below), only
the recent history, $t-s=O(t_{\mathrm b})$, affects the bath force at time $t$. Assuming that $\langle(v/v_s)^2\rangle$ is of order one,
the history displacement over this time is at most $O(\varepsilon q_c)$. The
perturbation \eqref{perturbapp} is then small compared with the medium dynamics
provided $\mu\lambda\,\varepsilon q_c\max|\mathcal{F}'|\ll v_0$. Thus
$\varepsilon\ll1$, together with weak coupling, provides a slow-probe regime in
which the linear-response expansion holds.

It should be emphasized, that these conditions describe a conservative regime of applicability,
rather than a sharp boundary of validity. The resulting weak-coupling description
may therefore remain accurate outside the slow-probe regime, provided
the coupling-weighted history perturbation remains small. Our
simulations also explore this broader regime, where the accuracy of
the reduced description is assessed by comparison with the full
microscopic dynamics.

\subsection*{Determining $t_b$}
To leading order in weak coupling, the bath relaxation scale can be
obtained at $\lambda=0$. 
Let $P_\pm(x,t)$ be the probability distributions of the RTPs with $\sigma=\pm1$. Defining $\rho(x)=P_+(x)+P_-(x)$ and $m(x)=P_+(x)-P_-(x),$ the Fokker--Planck equations are
\begin{align}
    \partial_t\rho&=-v_0\,\partial_xm+\mu\lambda q\,\partial_x\!\left[\mathcal F'(x)\rho\right],\\
    \partial_t m&=-v_0\,\partial_x\rho+\mu\lambda q\,\partial_x\!\left[\mathcal F'(x)m\right]-\frac{m}{\tau}.
    \label{eq:densitypolarity}
\end{align}
For $\lambda=0$, the two eigenvalues of the above equation are given by
\begin{equation}
 z_n^\pm=-\frac{1}{2\tau}
       \pm\sqrt{\frac{1}{4\tau^2}-\Omega_n^2}.
 \label{app:sep:eigenvalues}
\end{equation}
For $n=0$, $\rho$ is conserved, while for $n>0$, the slowest decay mode is $g_n=-\text{Re}[z_n^+]$, the smallest of these being $g_1$. To determine the relaxation time relevant to the probe, we look at Fourier coefficients of the contact profile,
\begin{equation}
 \mathcal F(x)
 =1+2\sum_{n\geq1}f_n\cos(k_nx),
 \qquad
 f_n=\frac{I_n(p)}{I_0(p)}.
\end{equation}
For $p>0$, $f_1>0$, so the force couples to the slowest nonuniform
mode. We therefore choose
\begin{equation}
 t_{\mathrm b}=\frac{1}{g_1}
 =\left[
 \frac{1}{2\tau}
 -\operatorname{Re}\sqrt{\frac{1}{4\tau^2}-\Omega_1^2}
 \right]^{-1}.
 \label{app:sep:gap}
\end{equation}
In the persistent regime, $\Omega_1\tau>1/2$, one has $t_{\mathrm b}=2\tau$, so that $\varepsilon=2\tau\omega_0$ is large for highly persistent media; there the comparison with the simulations decides.

\section{Effective medium forces}
\label{app1}
In this appendix, we derive the induced forces by the medium, as given in the main text.

\subsection{Mean force}
We consider the medium dynamics Eq.~\eqref{eq:eom-x} for a fixed position $q$ of the probe. Let $P_\pm(x)$ be the stationary probability distributions of the RTPs with $\sigma=\pm1$.
In the weak coupling limit, the stationary limit of the Fokker-Planck equations Eqs.~\eqref{eq:densitypolarity} admits a perturbative series $\rho(x)=\frac{1}{L}+\lambda \rho_1(x)$, and $m(x)=\lambda m_1(x)$. At the zeroth order,
\begin{align}
\rho_0(x)=\frac{1}{L},
\qquad
m_0(x)=0.
\end{align}
Putting this back we have at first order,
\begin{equation}
    v_0\,\partial_xm_1=\frac{\mu q}{L}\mathcal F''(x),\qquad
    v_0\,\partial_x\rho_1=-\frac{m_1}{\tau}.
\end{equation}
Using $\int_0^L\rho_1\,\dd x=0$,
\begin{equation}
    m_1(x)=\frac{\mu q}{Lv_0}\mathcal F'(x),\qquad
    \rho_1(x)=-\frac{\mu q}{Lv_0^2\tau}\left[\mathcal F(x)-\la\mathcal F\ra_0\right].
\end{equation}
Thus, the quasistatic force exerted by one particle becomes
\begin{align}
F_i^{\mathrm{qs}}(t)
&=
-\lambda \int_0^L dx\, \rho(x)\mathcal F(x)=
-\lambda \langle \mathcal F\rangle_0
+
\lambda^2 \frac{2\alpha\mu}{v_0^2} q(t)
\left(
\langle \mathcal F^2\rangle_0 - \langle \mathcal F\rangle_0^2
\right).
\label{eq:meanf}
\end{align}
Summed over the $N_b$ particles, this is Eq.~(\ref{eq:kappa}) with $\la\mathcal F\ra_0=1$. The induced stiffness $\kappa$ is that of a thermal particle bath at the effective temperature $T_{\text{RT}}=\zeta v_0^2\tau$. For the von Mises profile, one can obtain~\cite{ndc2026},
\begin{align}
 \la\mathcal F^2\ra_0-\la\mathcal F\ra_0^2=\frac{I_0(2p)}{I_0(p)^2}-1.  
\end{align}

\subsection{Friction}

To find the friction term, let us first note that, a perturbation by the probe modifies the forward generator of a particle by $q(t)L_1^\dagger$, with
\begin{align}
    L_1^\dagger P=\mu\lambda\,\partial_x\!\left[\mathcal F'(x)P\right].
\end{align}
Using the Kubo-Agarwal linear response formula for an observable $O$ as
\begin{equation}
    \la O(t)\,|\,q(s),\,s\le t\ra=\la O\ra_{q(t)}+\int_{-\infty}^t\dd s\,R_O(t-s)\,(q(s)-q(t)),
    \qquad
    R_O(s)=\left\la O(x(s))\,\frac{L_1^\dagger P_0}{P_0}(x(0))\right\ra_0 .
\end{equation}
The free density $P_0$ is uniform and independent of $\sigma$, so that $L_1^\dagger P_0/P_0=\mu\lambda\,\mathcal F''(x)$. For $O=F_A$ (see Eq.~\eqref{eq:FA}, 
\begin{align}
    R(t)=-\mu\lambda^2N_b\,\la\mathcal F(x(t))\,\mathcal F''(x(0))\ra_0 .
    \label{eq:Rdef}
\end{align}
Then, using Eq.~\eqref{eq:memory-kernel}, the friction kernel is
\begin{equation}
\Gamma(u) = \int_u^\infty\dd s\,R(s).    \label{eq:Gamma-def}
\end{equation}

To evaluate the correlation appearing in the response function Eq.~\eqref{eq:Rdef}, we rewrite 
\begin{align}
\mathcal F(x)=\sum_{n\in\mathbb Z} F_n e^{ik_n x},
\qquad\text{with  }
k_n=\frac{2\pi n}{L}.
\end{align}
 and $F_{-n}=F_n$, the average over the uniform initial position leaves only terms with opposite wavenumbers,
\begin{align}
    \la\mathcal F(x(t));\mathcal F(x(0))\ra_0=\sum_{n\neq0}F_n^2\,C_n(t),
    \qquad
    \la\mathcal F(x(t))\,\mathcal F''(x(0))\ra_0=-\sum_{n\neq0}k_n^2F_n^2\,C_n(t),
\end{align}
with the characteristic function of the displacement of a free particle,
\begin{align}
    C_n(t)=\la e^{\imu k_n[x(t)-x(0)]}\ra_0.
    \label{eq:cnapp}
\end{align}
which is just the generating function of an RTP without perturbation, which is well-known~\cite{dhar2019run}. Using this, we have, 
\begin{align}
\tilde R(\omega)=\mu\lambda^2\sum_{n\geq1}a_nk_n^2\,\frac{1-\imu\omega\tau}{\Omega_n^2\tau-\omega^2\tau-\imu\omega},
\end{align}
The friction kernel can then be obtained as
\begin{align}
    \tilde\Gamma(\omega)=\mu\lambda^2\sum_{n\geq1}a_nk_n^2\,
    \frac{1-\Omega_n^2\tau^2-\imu\omega\tau}{\Omega_n^2\left(\Omega_n^2\tau^2-\omega^2\tau^2-\imu\omega\tau\right)},
\end{align}
which is the result given in the main text.

\subsection{Noise}
The fluctuating part of the force is $\eta=F_A-\la F_A\,|\,q(s\le t)\ra$. It carries a factor $\lambda$, so to leading order its covariance is that of $F_A$ in the free medium,
\begin{align}
    \la\eta(t)\eta(t')\ra=\lambda^2N_b\la\mathcal F(x(t));\mathcal F(x(t'))\ra_0+O(\lambda^3)
    =\lambda^2\sum_{n\geq1}a_n\,C_n(|t-t'|)+O(\lambda^3).
\end{align}
Its spectrum, $S(\omega)=2\lambda^2\sum_na_n\,\mathrm{Re}\,\tilde C_n(\omega)$, follows from Eq.~\eqref{eq:cnapp},
\begin{align}
    S(\omega)=\lambda^2\sum_{n\geq1}\frac{2a_n\Omega_n^2\tau}{(\Omega_n^2-\omega^2)^2\tau^2+\omega^2}.
\end{align}
In the time-scale-separated regime, $\eta$ acts as a white noise,
\begin{align}
    \la\eta(t)\eta(t')\ra\simeq S(0)\,\delta(t-t'),\qquad
    S(0)=\lambda^2\frac{2}{\tau}\sum_{n\geq1}\frac{a_n}{\Omega_n^2}.
    \label{eq:noisecovfin}
\end{align}
\section{Numerical simulations}
We simulate Eqs.~\eqref{eq:chain-all} and \eqref{eq:eom-x} using velocity-Verlet updates for the chain, combined with explicit Euler updates for the medium particles and exact right-boundary damping. The tumbles occur independently with probability $(1-e^{-\Delta t/\tau})/2$ per step. We use $\Delta t=10^{-3}$ for $v_0<50$, and $\Delta t=10^{-4}$ for larger values of $v_0$. The stationary state averages are computed by letting the system relax to reach a stationary state, and thereafter time-averaging in the stationary state. 

\bibliographystyle{apsrev4-2}

\bibliography{refs-active}

\end{document}